\documentclass[twocolumn,showpacs,preprintnumbers,amsmath,amssymb,prb,a4paper]{revtex4}
\usepackage{amssymb,amsmath,amsbsy}

\usepackage{mathrsfs}
\usepackage{graphicx}
\usepackage{epstopdf}
\usepackage{dcolumn}
\usepackage{bm}
\usepackage{verbatim}
\usepackage{subfigure}
\usepackage{color}
\usepackage{braket}
\usepackage{braket}
\usepackage{multirow}  
\usepackage{array}
\usepackage{changepage}
\usepackage{float}

\newcommand{\ersample}{$^{167}\text{Er}\text{:Y}\text{SO}\ $ }
\newcommand{\er}{$^{167}\text{Er}^{3+} \ $}

\definecolor{green2}{rgb}{0,0.5,0}

\begin{document}
\title{Coupling erbium spins to a three-dimensional superconducting cavity \\
at zero magnetic field}

\author{Yu-Hui Chen}
\thanks{stephen.chen@otago.ac.nz}
\author{Xavier Fernandez-Gonzalvo}
\author{Jevon J. Longdell}
\thanks{jevon.longdell@otago.ac.nz}

\affiliation{The Dodd-Walls Centre for Photonic and Quantum Technologies \&  Department of Physics, University of Otago, 730 Cumberland Street, Dunedin, New Zealand.}

\date{\today}

\begin{abstract}
We experimentally demonstrate the coupling at zero magnetic field of an isotopically pure erbium doped yttrium orthosilicate crystal ($^{167}$Er:YSO) to a three-dimensional superconducting cavity with a $Q$ factor of $10^5$. A tunable loop-gap resonator is used, and its resonance frequency is tuned to observe the hyperfine transitions of the erbium sample. The observed spectrum differs from what is predicted by the published spin Hamiltonian parameters. The narrow cavity linewidth also enables the observation of asymmetric line shapes for these hyperfine transitions.  Such a broadly tunable superconducting cavity (from 1.6 GHz to 4.0 GHz in the current design) is a promising device for building hybrid quantum systems.
\end{abstract}

\maketitle

Quantum information systems rely on the performance of the transfer, storage, and processing of quantum states. Diverse platforms, such as  atoms, photons and macroscopic superconducting qubits \cite{Buluta2011,Georgescu2014,You2011}, have shown their own advantages in building different subsystems of a full quantum network. To combine the best features of these parts, such as the long coherence times of atoms, the ability of telecom photons to send quantum states over long distances and the flexibility of superconducting qubits, hybrid quantum systems have been proposed \cite{Kurizki2015,Xiang2013}. One important physical realization is to couple atomic spins to a microwave cavity \cite{Kubo2010a,Schuster2010,Probst2013}, where the electromagnetic field can be used to readout the spin states and the spins themselves can either be used to store or to process quantum information. Cavity quantum electrodynamics (cavity QED) provides the basic framework towards this goal.

Erbium doped crystals have attracted considerable interest in the field of hybrid quantum systems due to their optical transition near 1.5\,$\mu$m and their long optical coherence times \cite{Thiel2011}. It has been proposed by different groups to use them to perform quantum frequency conversion between microwave and optical photons \cite{Williamson2014,OBrien2014}. The first step towards this goal, i.e., the coherent conversion of microwave signals into optical ones, has recently been demonstrated experimentally \cite{Fernandez-Gonzalvo2015}. In addition, one of the stable isotopes of erbium, $^{167}$Er, has non-zero nuclear spin ($I$\,=\,7/2). $^{167}$Er:YSO crystals exhibit rich hyperfine level structures with energy splittings of a few GHz at zero magnetic field, and the prospect of using their nuclear spins to store quantum states is very attractive \cite{Wolfowicz2014}.

Several experimental results have been reported where erbium spins are coupled to various microwave cavities \cite{Probst2013,Tkalcec2014,Probst2014,Probst2014b}. Strong coupling has been demonstrated in two-dimensional coplanar superconducting cavities \cite{Probst2013,Tkalcec2014} as well as in three-dimensional (3D) copper cavities \cite{Probst2014}. These reports are based on magnetically tuning the spin transitions to match a fixed cavity frequency. In this paper we present the experimental results of a $^{167}$Er:YSO crystal coupled to a 3D superconducting niobium cavity at zero magnetic field. In our experiment the frequency of a tunable loop-gap resonator is scanned to match the zero field transitions of $^{167}$Er:YSO, which are detected as zero-field electron paramagnetic resonance (EPR) signals.

A high $Q$ factor cavity, a controllable environment to reduce decoherence effects, and a narrow inhomogeneous linewidth are desired for most cavity QED experiments. Compared to other systems our approach shows advantages in all three aspects. First of all, the superconductivity of the cavity is not deteriorated by an applied magnetic field \cite{Graaf2012}. At the same time, because the sample is placed at a minimum of the cavity's electric field, dielectric losses are minimal. These two facts allow our cavity to work with $Q$ factors of the order of $10^5$, giving a higher sensitivity that in turn allows us to see finer structures. Other than polishing it, no special surface treatment was carried out for our niobium cavity. Note that the state of the art 3D microwave cavities or Fabry-P$\acute{\text{e}}$rot cavities \cite{Kuhr2007} have $Q$ factors of $\mathcal{O}(10^{10})$. Secondly, a 3D cavity enables a controllable coupling of the qubits to their environment, which can be used to reduce decoherence\cite{Paik2011,Rigetti2012} and even to engineer the quantum bath seen by the qubits \cite{Murch2013}. And thirdly, narrow inhomogeneous linewidths are generally expected at zero magnetic field not only because of the absence of an imperfect magnetic field that would broaden the linewidths, but also because the absence of a magnetic field itself makes the magnetic inequivalence of atoms vanish. \cite{Bramley1983a}.

\begin{figure}
\includegraphics[width=0.48\textwidth]{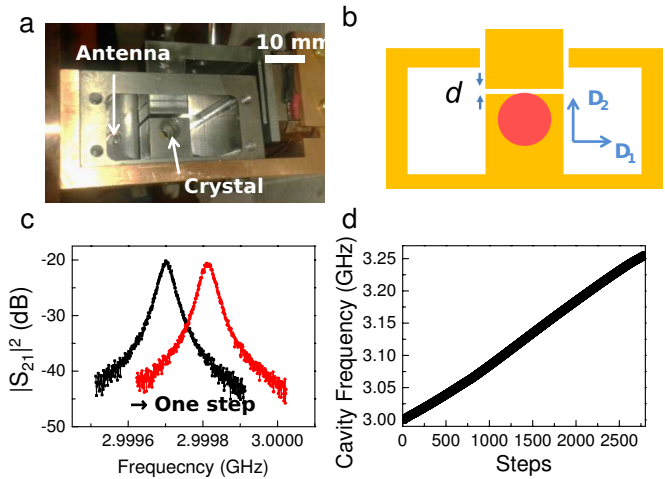}
\caption{(a) Photo of the loop-gap resonator. The $^{167}$Er:YSO crystal sits in the central hole. Two straight antennas are used to couple microwaves in and out. A separated plunger mounted on a piezo actuator is able to move in and out to change the gap size. (b) Schematic picture of the resonator. The resonant frequency can be shifted by changing the gap size $d$. The principal dielectric axes of the YSO crystal are oriented as noted. (c) Typical cavity transmission curves. We show two $|S_{21}|^2$ curves recorded at two consecutive steps of the piezo actuator. The typical linewidth of the cavity is 30\,kHz (d) Frequency tunability of the cavity as the plunger is moved step by step.}
\end{figure}

Our experimental setup is shown in Fig. 1(a). The erbium sample sits in the central hole of the loop-gap resonator. The used sample is a YSO cylindrical crystal of 4.95\,mm diameter and 12\,mm length, with 50 parts per million of the Y$^{3+}$ ions substituted by isotopically pure $^{167}$Er$^{3+}$ ions (Scientific Materials Inc). YSO has three orthogonal optical extinction axes $\mathbf{D_1}$, $\mathbf{D_2}$, and $\mathbf{b}$. In our sample the $\mathbf{b}$ axis is aligned along the long axis of the cylinder, and the $\mathbf{D_1}$-$\mathbf{D_2}$ plane is parallel to the end faces. The $\mathbf{D_2}$ axis is perpendicular to the resonator's gap, as illustrated in Fig 1(b). YSO has a monoclinic structure with two inequivalent sites (site 1 and site 2) where the $^{167}$Er$^{3+}$ ions can substitute yttrium ions. If an external static magnetic field is not aligned along the $\mathbf{b}$ axis or on the $\mathbf{D_1}$-$\mathbf{D_2}$ plane, one can see two magnetically inequivalent subclasses for each site \cite{Sun2008}.

The $^4I_{15/2}$ ground state of erbium  is split into eight levels by the crystal field of YSO, with only the lowest two levels being populated at cryogenic temperatures. Our cavity and sample assembly is cooled down to 5.1 K (measured at the body of the cavity) by the use of a home built vibration isolated cryostat (cooling head: Cryomech PT405). Because \er ions  have nuclear spin $I$\,=\,7/2 and an effective electronic spin $S$\,=\,1/2, the crystal field levels are split into 16 hyperfine levels even in the absence of an external magnetic field. These hyperfine splittings can be understood from the spin Hamiltonian \cite{Guillot-Noel2006}
\begin{equation}
H=\beta_e \mathbf{B} \cdot \mathbf{g} \cdot \mathbf{S} + \mathbf{S} \cdot \mathbf{A} \cdot \mathbf{I} +\mathbf{I} \cdot \mathbf{Q} \cdot \mathbf{I} -\beta_n g_n \mathbf{B} \cdot \mathbf{I} ,
\end{equation} 
where $\beta_e$ is the Bohr magneton, $\mathbf{B}$ the applied magnetic field, $\mathbf{g}$ the g-matrix of Zeeman effect, $\mathbf{A}$ the hyperfine matrix, $\mathbf{Q}$ the electric quadrupole matrix, $\beta_n$ the nuclear magneton, and $g_n=-0.1618$ is the nuclear $g$ factor. At 5.1 K all these 16 levels are populated and there are 120 possible transitions between them. Matching the cavity frequency to one of these transitions will give out a detectable zero-field EPR signal if the transition strength is high enough\cite{Bramley1983a}.

\begin{figure}
\includegraphics[width=0.5\textwidth]{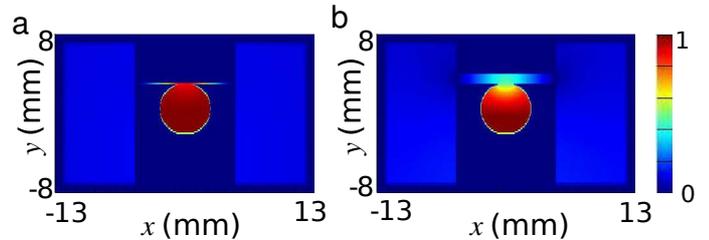}
\caption{Field distributions inside the tunable loop-gap resonator. (a) 0.1\,mm gap, corresponding to a cavity resonance frequency of 2.51\,GHz. (b) 1\,mm gap, giving a cavity frequency of 5.27\,GHz. The color scale shows the magnitude of the oscillating magnetic field $|B|$,  which is normalized to the individual maxima for each case.}
\end{figure}    

The design of our tunable cavity is based on a typical three-loop, two-gap configuration \cite{Wood1984}. The gaps here act as capacitors while the loops act as inductors, so the cavity resonance is analogous to that of an LC circuit. Shown in Fig.~1(b) is a schematic of the cavity. A plunger mounted on a piezo actuator can be moved in and out to change the gap $d$, which alters the capacitance of the equivalent LC circuit, allowing us to tune the resonance frequency of the cavity. The typical step size of the actuator is 30\,nm, which corresponds to a 0.1\,MHz frequency shift. The setup shown in Fig.~1(a) has a tunable range from 1.6\,GHz to 4.0\,GHz. To minimize the energy radiated out from the cavity, a cap is added on each end. Two short straight wires serving as coupling antennas are inserted into the cavity through a 5\,mm diameter window drilled on each cap. The two antennas are connected to a network analyser in order to measure the transmission signal $|S_{21}|^2$. Attenuators and amplifiers are used at room temperature but their contributions in $|S_{21}|^2$ are already subtracted in all the figures shown in this paper (although not the signal losses in the coaxial cables inside the cryostat).

By scanning the input microwave frequency, we can find the resonance frequency of the cavity as a transmission peak, as shown in Fig. 1(c). Our cavity has a typical linewidth of 30\,kHz (when it is off resonance with an erbium transition), corresponding to a quality factor $Q \approx 9 \times 10^4$. This linewidth is not the intrinsic linewidth of the cavity, as it is broadened by the vibrations of the cryostat's cooling head. A change of merely 0.3\,nm in the gap size is equivalent to a resonance frequency shift of 1\,kHz, making our measurements quite sensitive to vibrations. Our home built cryostat is designed to be able to keep the sample cold for more than an hour after being switched off. Small vibrations are still present even when the cooling head is off, but they are reduced immensely. Under these conditions we were able to measure $Q \approx 2 \times 10^5$. Figure 1(d) confirms the tunability of our cavity: as the plunger is moved step by step, the cavity resonant frequency goes up almost linearly.

The loop-gap resonator concentrates the electromagnetic field inside the central hole where the sample sits, therefore obtaining a good filling factor\cite{Wood1984,Williamson2014}. Besides that, another outstanding feature of our tunable cavity is that the field distribution in the central hole stays uniform during the cavity frequency tuning. In Fig.~2, we show the simulated (FDTD Solutions, Lumerical Solutions Inc.) field distribution for two different gap distances. These are cross sections in the middle of the cavity as viewed along the long axis of the crystal. When the gap $d=0.1$\,mm, the cavity frequency is 2.51\,GHz. It is clear from the simulation that most of the energy is confined uniformly in the central hole. When the gap is increased to $d=1$\,mm, the frequency goes up to 5.27\,GHz. The field distribution is still reasonably uniform, although some field leaks out from the central hole. In our zero-field EPR measurements, the frequency is scanned from 3.02\,GHz to 3.20\,GHz, and the field distribution can be considered to be the same throughout this range.

\begin{figure}
\includegraphics[width=0.48\textwidth]{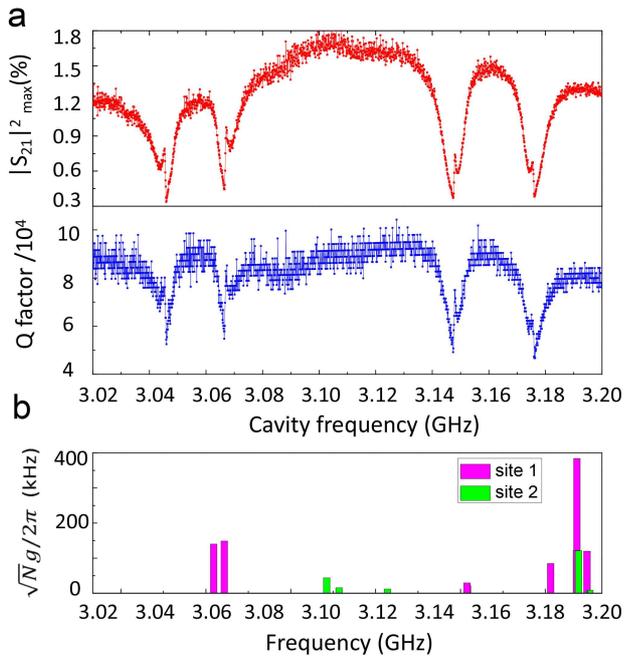}
\caption{(a) Experimental results of zero-field EPR. \emph{Top:} peak transmission as a function of cavity frequency (linear scale). \emph{Bottom:} $Q$ factor as a function of cavity frequency. (b) Calculated collective coupling strengths around 3.1\,GHz at zero magnetic field. The pink bars show transitions related to site 1, while the green bars show those related to site 2.}
\end{figure}

Figure 3(a) shows our experimental results when scanning the cavity from 3.02\,GHz to 3.20\,GHz. At each step of the actuator, we used the network analyser to measure the transmission spectrum of the resonator. Then the frequency for which $|S_{21}|^2$ is maximum is considered as the cavity's resonant frequency $\omega_c/2\pi$, with the corresponding transmission being $|S_{21}|^2_{\text{max}}$. When $\omega_c/2\pi$ is tuned on resonance with one of the hyperfine transitions of $^{167}$Er:YSO, a decrease in the transmission is expected due to the coupling between the cavity mode and the spin ensemble. In the weak coupling regime, the spin ensemble provides an additional loss channel for the cavity field and therefore both $|S_{21}|^2_{\text{max}}$ and the $Q$ factor drop near resonance, as can be observed in Fig. 3(a). We could identify four transmission minima within the explored range, at 3045.9\,MHz, 3066.5\,MHz, 3147.4\,MHz and 3176.2\,MHz, with widths of a few MHz. Note that the erbium ions have the effect of pulling the cavity frequency when they are close to resonances with the cavity. However, the frequency shifts in our experiment are on the order of 1\,kHz, therefore we neglected this effect in our data analysis.

\begin{figure}
\includegraphics[width=0.48\textwidth]{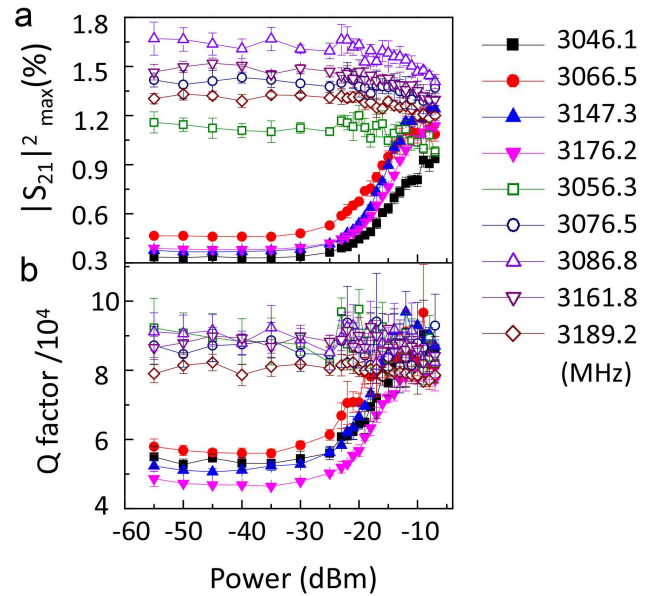}
\caption{Saturation of the hyperfine transitions at high input powers. (a) Peak transmission as a function of input power at several cavity frequencies. (b) $Q$ factor as a function of input power at several cavity frequencies. Here the input power is defined as the power immediately after the attenuators. The four solid symbol curves in each figure correspond to the four dips in Fig. 3, while the open symbol curves are taken at randomly chosen off-resonant frequencies.}
\end{figure}

To further confirm that these zero-field EPR signals come from the erbium atoms in the sample, we studied the behaviours at different input powers, as shown in Fig. 4. Four of the curves correspond to the four minima in Fig. 3(a), and the other five are measured at randomly chosen off-resonant frequencies. For each curve $|S_{21}|^2_{\text{max}}$ and the corresponding $Q$ factor are measured as a function of the input power. For the off-resonant frequencies the responses of $|S_{21}|^2_{\text{max}}$ and $Q$ factor show little dependence on the input power (open symbol curves in Fig. 4). We attribute the slight declining of these curves to the power dependency of superconductivity \cite{Graaf2012}. If $\omega_c/2\pi$ is tuned to match the spin transitions, the input microwaves are absorbed, which results in a lower $|S_{21}|^2_{\text{max}}$ and a lower $Q$ factor. When more and more photons go into the cavity, the erbium atoms are driven harder and harder and are eventually saturated. So both the transmission and the $Q$ factor increase, as can be seen in our measurements (solid symbol curves in Fig. 4). The turning points for the four dips are approximately at -25\,dBm, which corresponds to a total photon number inside the cavity of $5 \times 10^{12}$. At 5.1 K, the population number difference between two levels 3\,GHz apart in our sample is approximately $1 \times 10^{14}$. These numbers, together with the cavity mode volume obtained from the simulations and the parameters of the spin Hamiltonian \cite{Guillot-Noel2006} in Eq.~(1) enable us to estimate the driving Rabi frequency for the collective states of the erbium atoms, which is approximately 0.1\,MHz and approaches their $\sim$MHz inhomogeneous linewidths. The data from our experiments, unless otherwise noted, are all measured at -60\,dBm input power. This input power is low enough that the response from the erbium ions is linear, but the intra-cavity field still contains many photons.

The error bars in Fig. 4 show the contribution of vibrations in the background measurements. Vibrations can shift the cavity transmission curves, so the measured $|S_{21}|$ curve is therefore an average of several shifted curves, which is broader than the cavity's intrinsic curve. As the cavity linewidth is broadened out by the atoms, the vibration effects become less significant. This explains why the points with lower $Q$s in Fig. 4 have smaller error bars.

Based on Eq.~(1) and the matrices given in the literature\cite{Guillot-Noel2006}, we can calculate the transition frequencies of the erbium ions and their collective coupling strength to the loop-gap resonator $\sqrt{N}g$. The results for zero magnetic field and at a temperature of 5.1\,K are shown in Fig. 3(b). In these calculations, the effects of finite temperature are simply included by assuming a Boltzmann distribution of the atomic population in the different energy states. 

As shown in Fig. 3, the deviations between our measurements and the predicted transition frequencies are of the order of several tens of MHz. This shows that the spin Hamiltonian parameters determined at high magnetic fields \cite{Guillot-Noel2006} are imperfect when the results are extrapolated to zero magnetic field. A possible explanation for this are that the errors associated with the magnetic field and the sample orientation, which get carried into the calculated parameters during data fitting, and small errors can be seen when interpolating from high magnetic fields to zero field.

Because the linewidth of the superconducting cavity is approximately 30\,kHz, which is much less than the inhomogeneous linewidths of the erbium ions ($\mathcal{O}$(MHz)), it is possible to observe the line shapes of the hyperfine transitions.
 In Fig. 3(a) the transmission minima show asymmetric line shapes with a sharp edge. The inhomogeneous linewdiths are also narrower than what has been observed in Er:YSO by others \cite{Probst2013,Probst2014}.

We believe that both of these effects are a result of how the transition frequencies vary with magnetic field around zero magnetic field. A very crude way of modelling the zero field splittings is to use a hyperfine term in the spin Hamiltonian like $\alpha S_z I_z$. In such a situation the energy levels are joint eigenstates of the $z$-component of the electron ($S_z$) and nuclear ($I_z$) angular momentum and come as pairs of doubly degenerate levels. The additional terms in the hyperfine coupling and the nuclear quadrupole term cause mixing between these levels. This means that rather than crossing at zero magnetic field the energy levels display avoided crossings, which in turn means that at zero magnetic field all the transition frequencies have zero first-order Zeeman shift. The same situation has been behind a number of groundbreaking advances in the extension of coherence times \cite{McAuslan2012,Zhong2015}. When the inhomogeneous broadening  is considered as the perturbation of local magnetic fields, this zero first-order Zeeman shift could explain the narrower inhomogeneous linewidths that we observe as well as the asymmetric lineshapes (see Fig.5). The situation would be much clearer if we had spin Hamiltonian parameters that were more accurate at low magnetic fields.

In our experiment, we made no effort to shield the niobium cavity from the earth's magnetic field\cite{Aull2012} (approximately 100 $\mu T$). However, this effect can be ruled out because no changes in the transition frequencies or the line shapes were observed over different runs, where the cavity and sample were in different orientations.

 
\begin{figure}
\includegraphics[width=0.48\textwidth]{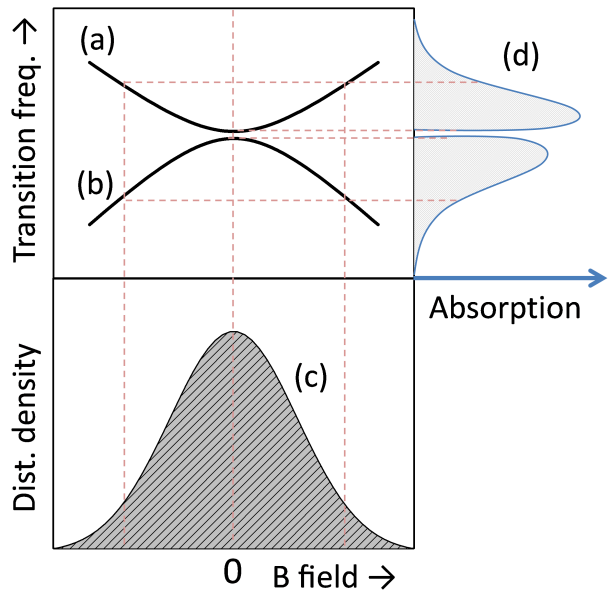}
\caption{A figure illustrating the mechanism whereby anti-crossing energy levels can lead to narrow lines as well as the asymmetric doublets seen in our experiments. Curves (a) and (b) show how the transition frequency varies with magnetic field. Curve (c) represents the distribution of random magnetic fields experienced by the erbium atoms due to other spins in the host. Curve (d) shows the resulting inhomogeneous line shapes.}
\end{figure} 

We can estimate from Fig. 3 that the linewidths of the hyperfine transitions are approximately $\Gamma^* /2 \pi = 5$\,MHz. An ensemble of spins inside a cavity is usually modelled as a harmonic oscillator coupled to a cavity mode\cite{Probst2013,Tkalcec2014}. Based on this model, we can calculate the collective coupling stength $\sqrt{N} g /2\pi = 150$\,kHz and the cooperativity factor $Ng^2/\kappa \Gamma^* \approx 0.2$, with $\kappa/2 \pi = 30$\,kHz being the cavity loss rate (it is worth noting that the asymmetric line shapes here hinder both the measurements of linewidths and the using of this model\cite{Diniz2011}). So we are not working in the strong coupling regime. Compared to the literature \cite{Probst2013}, where a coupling strength of 34\,MHz and a transition linewidth of 12\,MHz have been reported, the $\sqrt{N}g$ in our experiment is approximately $1/200$ of that. One way to achieve higher coupling strengths is to reduce the working temperature from the current 5.1\,K to mK range. This would increase the population differences between the involved energy levels by a factor of up to 400 for transitions starting from the lowest level, with a corresponding increase in coupling strengths. Once the spin Hamiltonian parameters are determined more accurately at zero magnetic field we may also benefit from looking for transitions of higher transition strengths. In addition, another factor that makes working at zero magnetic field appealing is that the hyperfine energy levels have zero first-order Zeeman shift and are non-sensitive to magnetic field fluctuations, therefore they may exhibit very long coherence times \cite{McAuslan2012}. 


In conclusion, we have demonstrated coupling of an \ersample crystal to a tunable 3D superconducting cavity at zero magnetic field at 5.1\,K. Hyperfine resonance lines are observed with collective coupling strengths to the cavity mode estimated to be approximately 150\,kHz. The observed spectrum differs from what is predicted by the published spin Hamiltonian parameters. A superconducting cavity with a high $Q$ factor has also enabled a detailed observation of the line shapes of the hyperfine transitions. Changing cavity frequencies electromagnetically with high resolution \cite{Wilson2011} has many advantages and could be achieved by, e.g., changing the gap size or coupling the cavity to superconducting qubits. Due to the narrow cavity linewidth, complicated manipulations can be carried out before photons leak out from the cavity, which shows prospective implementations in the field of quantum information. 

This work was sponsored by the Marsden Fund (Contract No. UOO1221) of the Royal Society of New Zealand.

\end{document}